\documentclass{article}

\def\xxinput#1{\input#1}

\xxinput{vsolj02.sty}

\usepackage[dvipdfmx]{graphicx}
\usepackage[comma,colon]{natbib}

\usepackage[OT2,T1]{fontenc}

\def\cite{\citealt}
\setcitestyle{aysep={}}

\newcounter{author}
\def\altaffilmark#1{$^{#1}$}
\def\altaffiltext#1{$^{#1}$\,}

\def\authorcount#1#2{{\refstepcounter{author}\label{#1}
                     \altaffiltext{\ref{#1}}{#2}}}

\begin{document}

\begin{center}

\title{Romanov V48: unusual intermediate polar in the period gap}

\author{
        Taichi~Kato\altaffilmark{\ref{affil:Kyoto}},
}
\email{tkato@kusastro.kyoto-u.ac.jp}
\author{
        Filipp~D.~Romanov\altaffilmark{\ref{affil:Romanov}}
}
\centerline{\textit{filipp.romanov.27.04.1997@gmail.com}, https://orcid.org/0000-0002-5268-7735} \vskip 0.2cm

\authorcount{affil:Kyoto}{
     Department of Astronomy, Kyoto University, Sakyo-ku,
     Kyoto 606-8502, Japan}

\authorcount{affil:Romanov}{
     Pobedy street, house 7, flat 60, Yuzhno-Morskoy, Nakhodka, Primorsky
     Krai 692954, Russia}

\end{center}

\begin{abstract}
\xxinput{abst.inc}
\end{abstract}

  The object Gaia EDR3 2236896418906579072 at
20$^{\rm h}$ 11$^{\rm m}$ 16\hbox{$.\mkern-4mu^{\rm s}$}825,
$+$60$^{\circ}$ 04$^\prime$ 28\hbox{$.\mkern-4mu^{\prime\prime}$}04
(J2000.0) has $BP$=17.407, $RP$=17.000 and a parallax
$\varpi$=0.693(65) mas \citep{GaiaEDR3}.
The variability of this object was detected by
the Asteroid Terrestrial-impact Last Alert System (ATLAS:
\cite{ATLAS}) as ATO J302.8201$+$60.0744 with a classification
of ``dubious'' (unspecified variable stars not classified
into the 12 categories including pulsating, eclipsing or other
representative variables) \citep{hei18ATLASvar}.
\citet{hei18ATLASvar} gave a period of 0.081328~d
(by Lomb-Scargle method: \cite{LombScargle}).
This object has an X-ray counterpart of 1RXS J201117.9$+$600421.
The Zwicky Transient Facility (ZTF: \cite{ZTF}) also
detected the variability of this object and listed it
as a candidate variable star with a range of 16.783--18.188~mag
and a possible period of 0.0404~d \citep{ofe20ZTFvar}.
Using Gaia DR2 \citep{GaiaDR2}, \citet{mow21Gaiavar} selected
candidate variable stars and listed this object as
a short time-scale candidate (DR2\_STS).  The catalog by
\citet{mow21Gaiavar} was released in 2020 September in arXiv.

  One of the authors (Filipp Romanov, hereafter FR) noticed
this object when comparing X-ray sources (such as ROSAT sources)
and ultraviolet-excess objects (GALEX: \cite{GALEX})
with blue stars on the Aladin Sky Atlas to search for
cataclysmic variables (CVs).
FR was the first to correctly identify
the period of 0.0420991~d using ZTF DR4 and the Pan-STARRS1
(PS1: \cite{PS1}) surveys.  Sebasti\'an Otero,
the chief supervisor of the American Association of Variable Stars
(AAVSO) Variable Star Index (VSX: \cite{wat06VSX})
suggested a classification of a candidate polar (AM:).
FR reported these results to the AAVSO VSX
with the name of Romanov V48 on 2021 February 23\footnote{
  $<$https://www.aavso.org/vsx/index.php?view=detail.top\&oid=2215537$>$.
}.
We hereafter use this designation for this object.

\begin{figure*}
\begin{center}
\includegraphics[width=16cm]{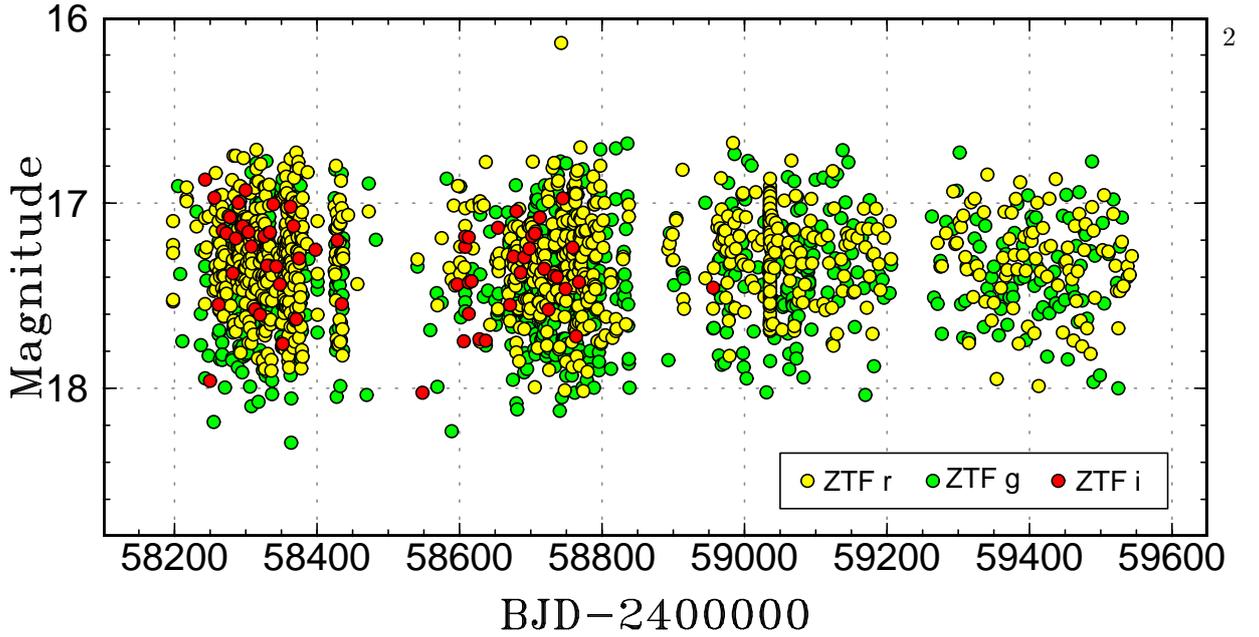}
\caption{
  Long-term ZTF light curve of Romanov V48.
}
\label{fig:long}
\end{center}
\end{figure*}

\begin{figure*}
\begin{center}
\includegraphics[width=16cm]{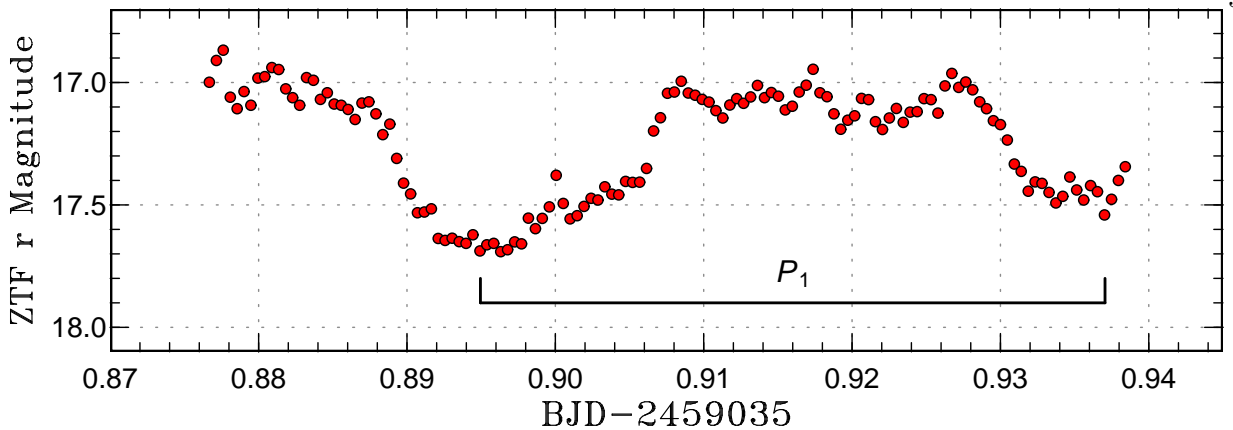}
\caption{
  ZTF light curve of Romanov V48 on 2020 July 5.
  Flickering and large-amplitude orbital variation
  are present.
}
\label{fig:lc}
\end{center}
\end{figure*}

\begin{figure*}
\begin{center}
\includegraphics[width=16cm]{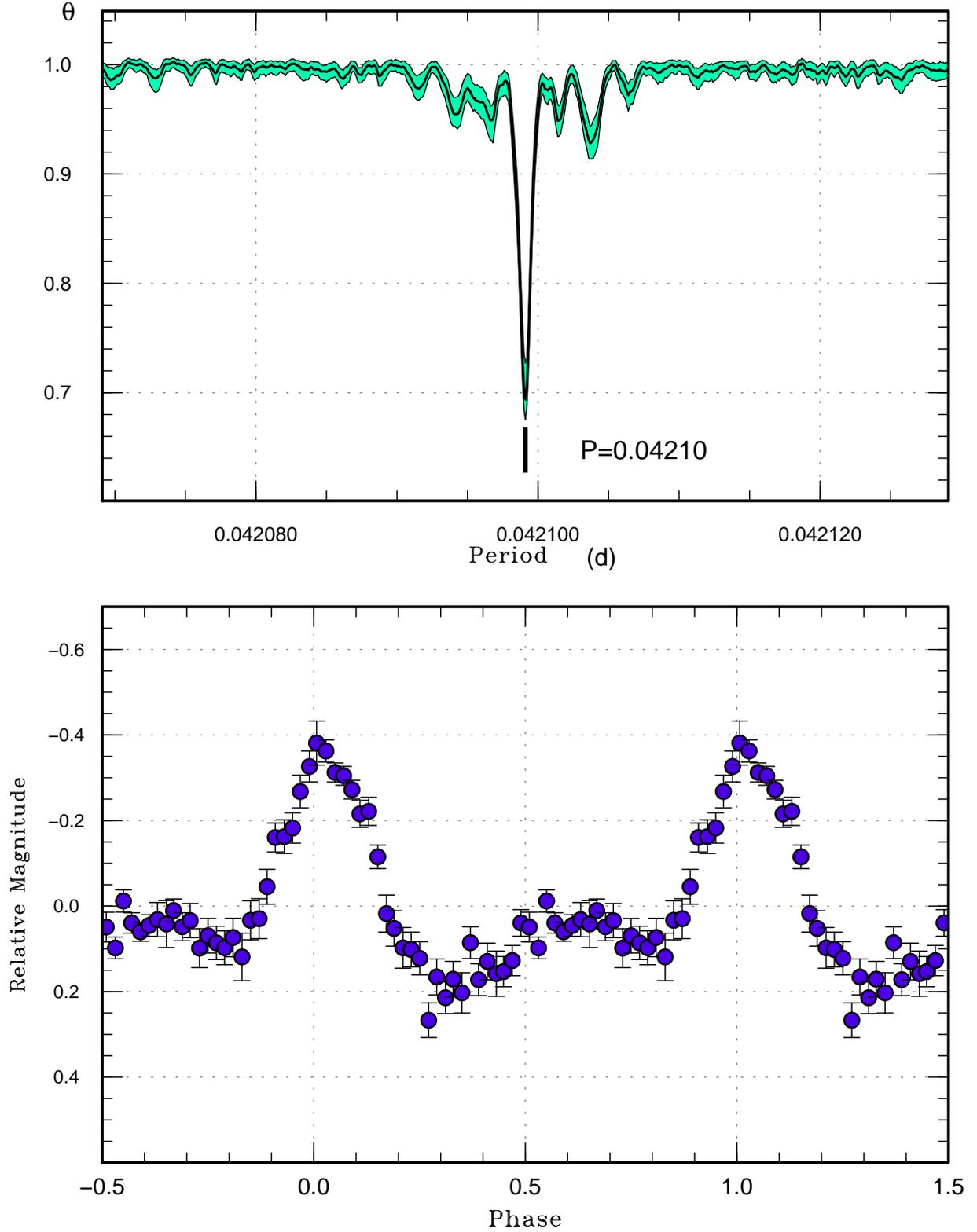}
\caption{PDM analysis of Romanov V48 using the ZTF data
  of the shorter period ($P_1$) = spin period.
  (Upper): PDM analysis.  The bootstrap result using
  randomly contain 50\% of observations is shown as
  a form of 90\% confidence intervals in the resultant 
  $\theta$ statistics.
  (Lower): mean profile.  1$\sigma$ error bars are shown.
}
\label{fig:pdm}
\end{center}
\end{figure*}

\begin{figure*}
\begin{center}
\includegraphics[width=16cm]{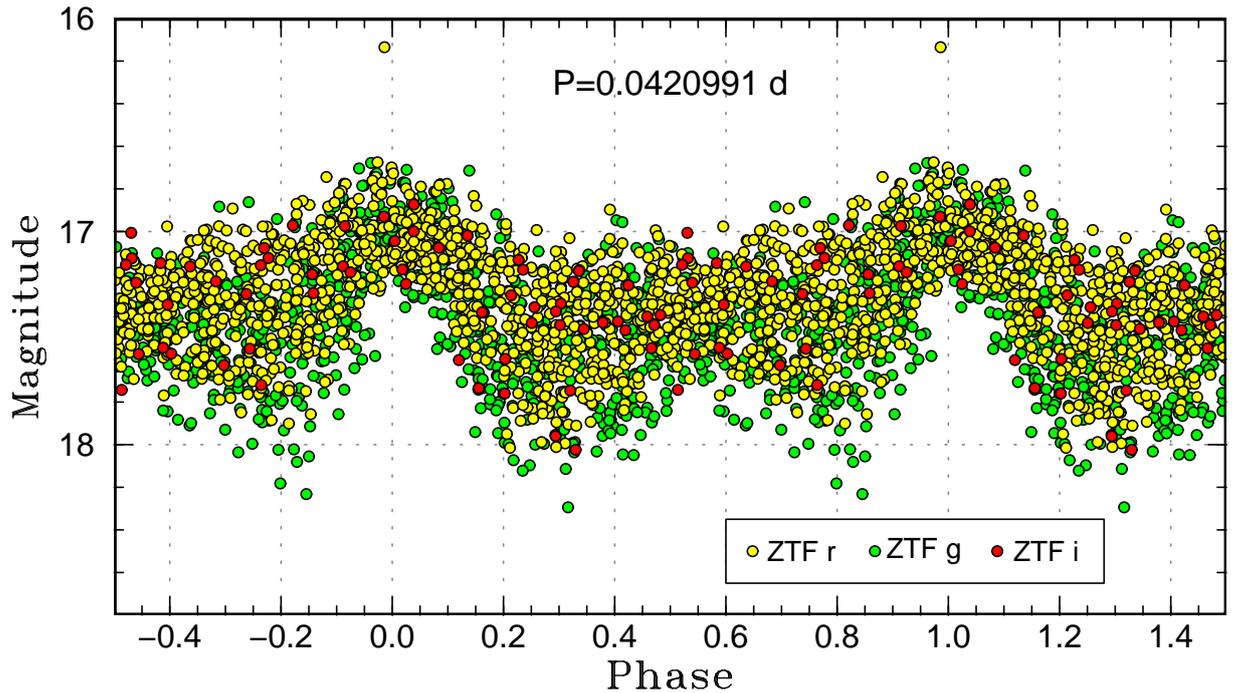}
\caption{The light curve of Romanov V48 using
  the ZTF data folded by $P_1$.
  The zero epoch was chosen as BJD 2458718.664.
}
\label{fig:phase}
\end{center}
\end{figure*}

  Since the period is far shorter than the ``period minimum''
of CVs \citep{kol93CVpopulation,how97periodminimum,
gan09SDSSCVs,kni06CVsecondary,kni11CVdonor,kat22stageA},
we made a more detailed analysis and tried to clarify
the nature of the object.
We analyzed the light variation using the ZTF public data\footnote{
  The ZTF data can be obtained from IRSA
$<$https://irsa.ipac.caltech.edu/Missions/ztf.html$>$
using the interface
$<$https://irsa.ipac.caltech.edu/docs/program\_interface/ztf\_api.html$>$
or using a wrapper of the above IRSA API
$<$https://github.com/MickaelRigault/ztfquery$>$.
}.
The long-term variation is shown in figure \ref{fig:long}.
There was no difference between $g$, $r$ and $i$
light curves and there was no high/low states which
are usually seen in many polars.
There was one sequence of time-resolved photometry
in the ZTF data (figure \ref{fig:lc}).  The presence
of flickering is characteristic to a CV.
The box-shaped, large-amplitude (0.7~mag)
orbital variation is similar to that of a polar and
does not resemble the orbital variation of a dwarf nova.
A phase dispersion minimization (PDM: \cite{PDM}) analysis
of the ZTF data confirmed the period detected by FR
(figure \ref{fig:pdm}).  The period was determined
to be 0.04209909(2)~d, whose error was estimated by
the methods of \citet{fer89error} and \citet{Pdot2}.
The light curve phased with this period is shown in
figure \ref{fig:phase}.  There is no difference
in the profile between the three ZTF bands and
the colors were almost zero.  The presence of two
maxima of different amplitudes within one phase
excludes the possibility of the double period being
the true one.

\begin{figure*}
\begin{center}
\includegraphics[width=16cm]{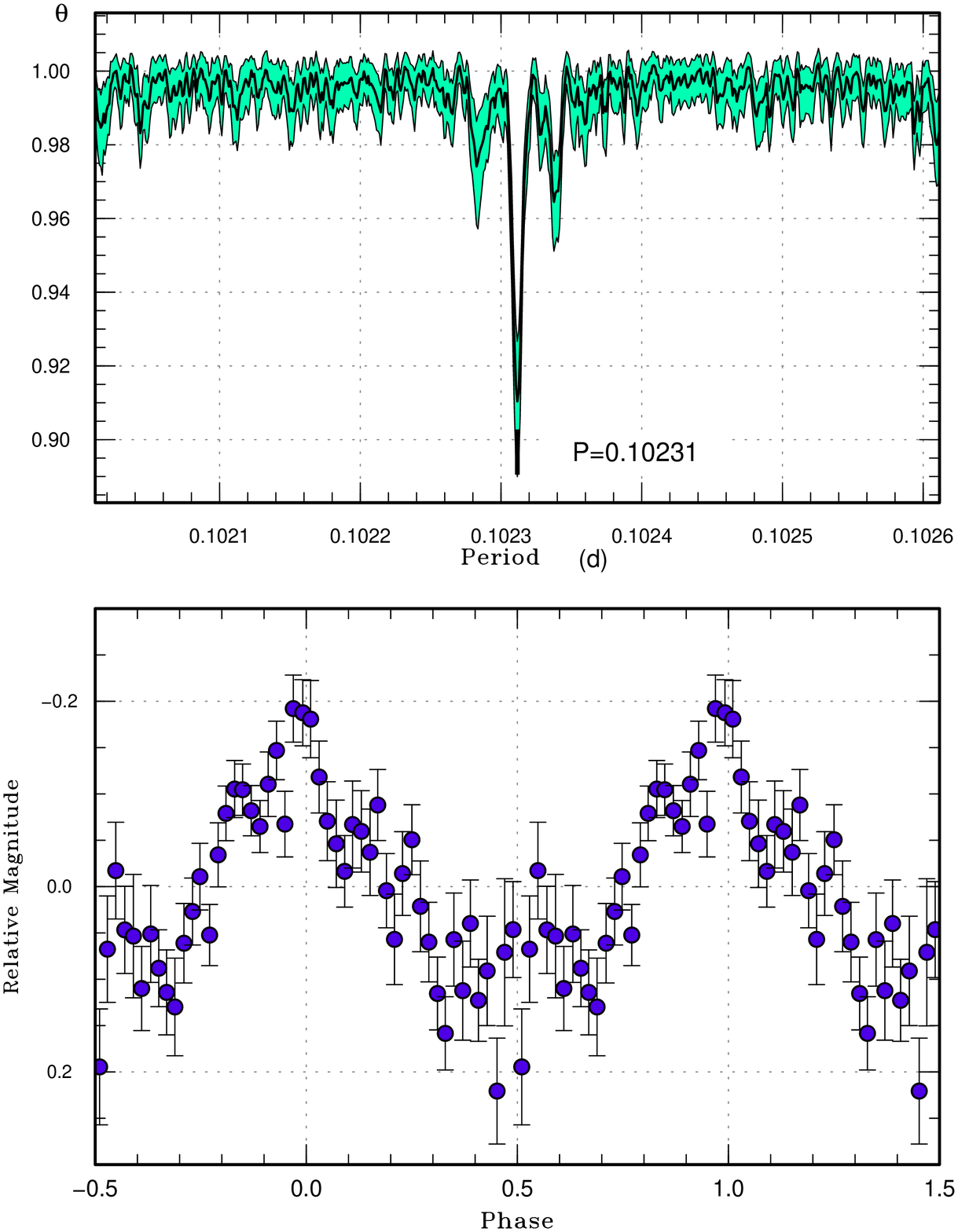}
\caption{PDM analysis of the longer period ($P_2$)
  = orbital period of Romanov V48 using the ZTF data.
  (Upper): PDM analysis.
  (Lower): mean profile.
}
\label{fig:pdm2}
\end{center}
\end{figure*}

   \citet{bar21tesssdBs} reported on a search for pulsating
subdwarf B star using the Transiting Exoplanet Survey
Satellite (TESS) full frame images and Romanov V48
was included in their sample (under the name of
2236896418906579072).  \citet{bar21tesssdBs} included
this object among 30 sdBV candidates that were not
spectroscopically classified (in their table 3
and figure 7).  Although the power spectrum by
\citet{bar21tesssdBs} detected the period
we determined at 275$\mu$Hz, \citet{bar21tesssdBs}
did not mention it.
The signal at 113$\mu$Hz in \citet{bar21tesssdBs}
is indeed present in the ZTF data.  The period has been
determined to be 0.1023116(4)~d (figure \ref{fig:pdm2}).
The amplitude of the 0.1023116-d period was 0.4~mag,
smaller than that of the 0.04209909-d period.
We could not detect a signal around 12$\mu$Hz (=0.96~d)
in the ZTF data.  We refer to the 0.04209909-d period as $P_1$
0.1023116-d period one as $P_2$.
The maxima can be expressed as
\begin{equation}
{\rm Max} \;(P_1)\; ({\rm BJD}) = 2458718.664(1) + 0.04209909(2) E
\label{equ:eph1}
\end{equation}
and
\begin{equation}
{\rm Max} \;(P_2)\; ({\rm BJD}) = 2458718.657(1) + 0.1023116(4) E.
\label{equ:eph2}
\end{equation}

\begin{figure*}
\begin{center}
\includegraphics[width=16cm]{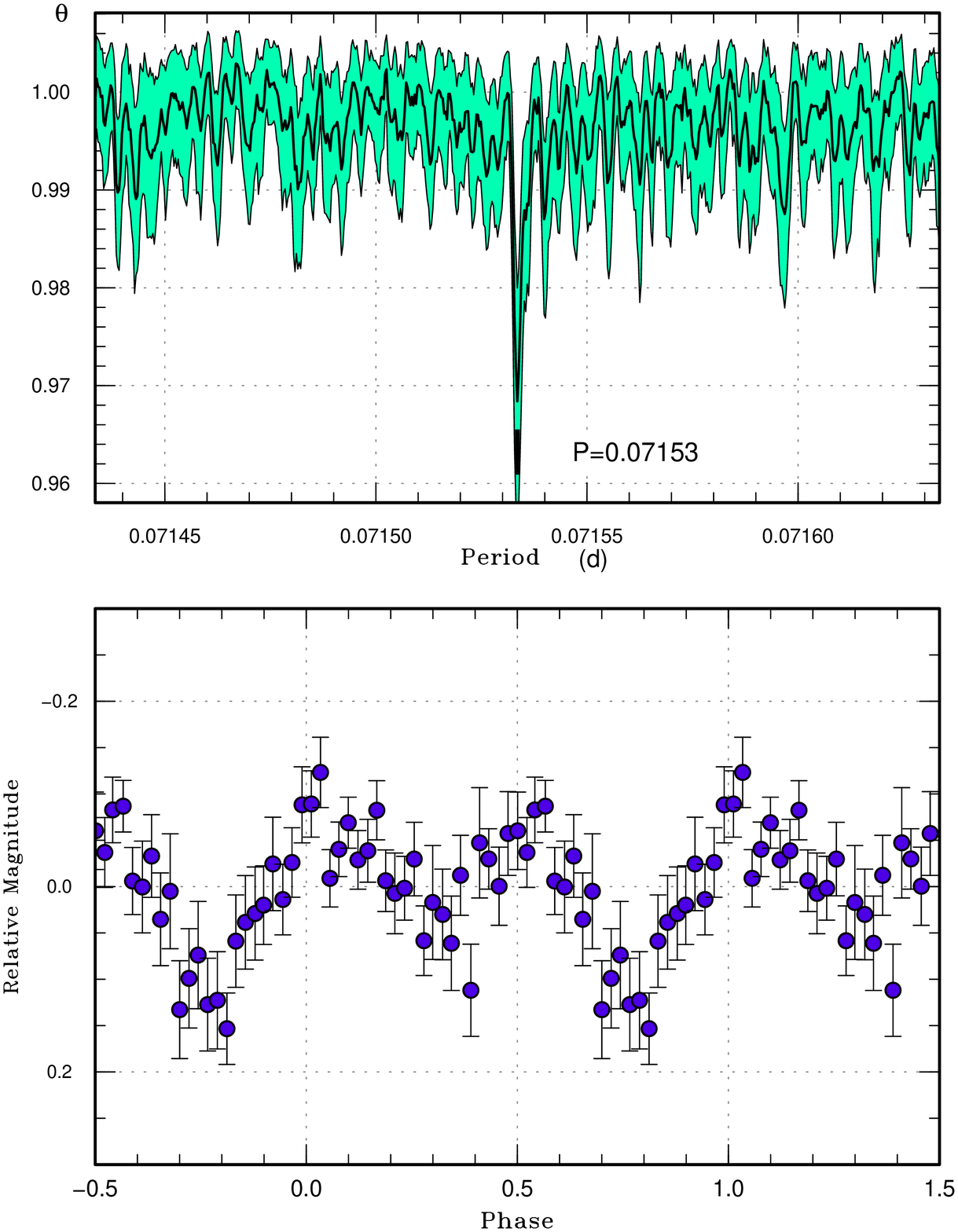}
\caption{PDM analysis of the beat period ($P_3$)
  of Romanov V48 using the ZTF data.
  (Upper): PDM analysis.
  (Lower): mean profile.
}
\label{fig:pdm3}
\end{center}
\end{figure*}

\begin{figure*}
\begin{center}
\includegraphics[width=15cm]{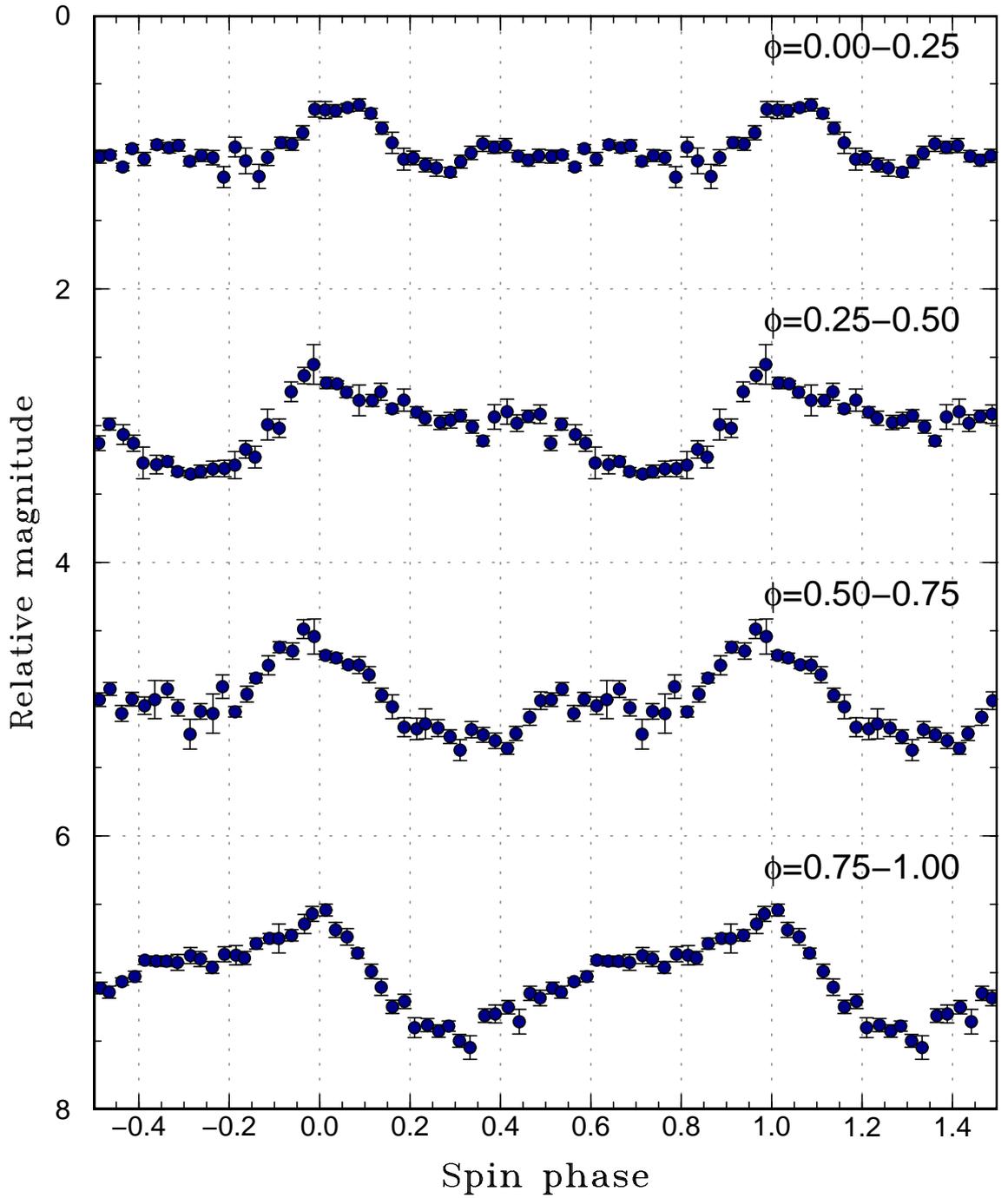}
\caption{Variation of pulse profiles of Romanov V48
  depending on the beat phase ($\phi$).  The beat phases
  $\phi$ were determined using the equation (\ref{equ:eph3}).
  The pulse phases were determined using the equation
  (\ref{equ:eph1}).  All bands of the ZTF data were
  combined.
}
\label{fig:phase2}
\end{center}
\end{figure*}

   The presence of two periods $P_1$ and $P_2$ suggests
an intermediate polar (IP).
In this interpretation, $P_2$ is the orbital period
($P_{\rm orb}$) and it is in the period gap.
In this case, we can expect a signal at the beat period
between $P_1$ (spin period) and $P_2$.  This is indeed the case
and a period $P_3$=0.0715335(2)~d was detected
(figure \ref{fig:pdm3}).  This signal was not apparent in
the figure of \citet{bar21tesssdBs}.  The ephemeris is
\begin{equation}
{\rm Max} \;(P_3)\; ({\rm BJD}) = 2458718.668(1) + 0.0715335(2) E.
\label{equ:eph3}
\end{equation}
This period reflects the period in which the beamed light
from the magnetic poles on the white dwarf sweeps
the surface of the secondary.  The existence of this
period is a strong support to the IP interpretation.
We could detect systematic variations of pulse profiles
depending on the beat phase (figure \ref{fig:phase2}).
These variations reflect the geometry of the magnetic
poles against the accretion stream from the secondary.

   IPs inside or below the period gap are
relatively rare and well-established ones having
$P_{\rm orb}$ shorter than Romanov V48 are:
V455 And \citep{ara05v455and},
HT Cam \citep{tov98htcam,kem02htcam},
V1025 Cen \citep{buc98v1025cen,hel02v1025cen}, 
DW Cnc \citep{pat04dwcnc},
EX Hya \citep{ste83exhya,jab85exhya}, 
CC Scl \citep{wou12ccscl,kat15ccscl},
AX J1853.3$-$0128, IGR J18173$-$\hspace{0pt}2509 \citep{tho13XrayCVs} and
Swift J0503.7$-$2819 \citep{hal15SwiftCVs}.
According to Koji Mukai's ``The Intermediate Polars''
page\footnote{
  $<$https://asd.gsfc.nasa.gov/Koji.Mukai/iphome/iphome.html$>$.
}, only one IP was plotted on his diagram
[V515 And: \citep{but08v515and,koz12v515and}; $P_{\rm orb}$
has not been directly determined].
Many IPs follow the relation between spin period
($P_{\rm spin}$) and $P_{\rm orb}$ on the line
$P_{\rm spin}/P_{\rm orb} \simeq 0.1$ or less
\citep{pat94IPreview}.
Among the objects listed, V1025 Cen, DW Cnc, EX Hya and
IGR J18173$-$2509 are the exceptions.
\citet{nor08IPaccretion} explained a larger fraction
of IPs having large $P_{\rm spin}/P_{\rm orb}$ in
short-$P_{\rm orb}$ systems considering the evolution
and the spin equilibrium.
Romanov V48 appears to be most similar to DW Cnc
although Romanov V48 is in the period gap and has
larger amplitudes of spin pluses.  The absolute magnitude
of Romanov V48 ($M_V$=$+$6.5) is the brightest
among these objects except IGR J18173$-$2509 whose distance
is unknown.

\begin{figure*}
\begin{center}
\includegraphics[width=13cm]{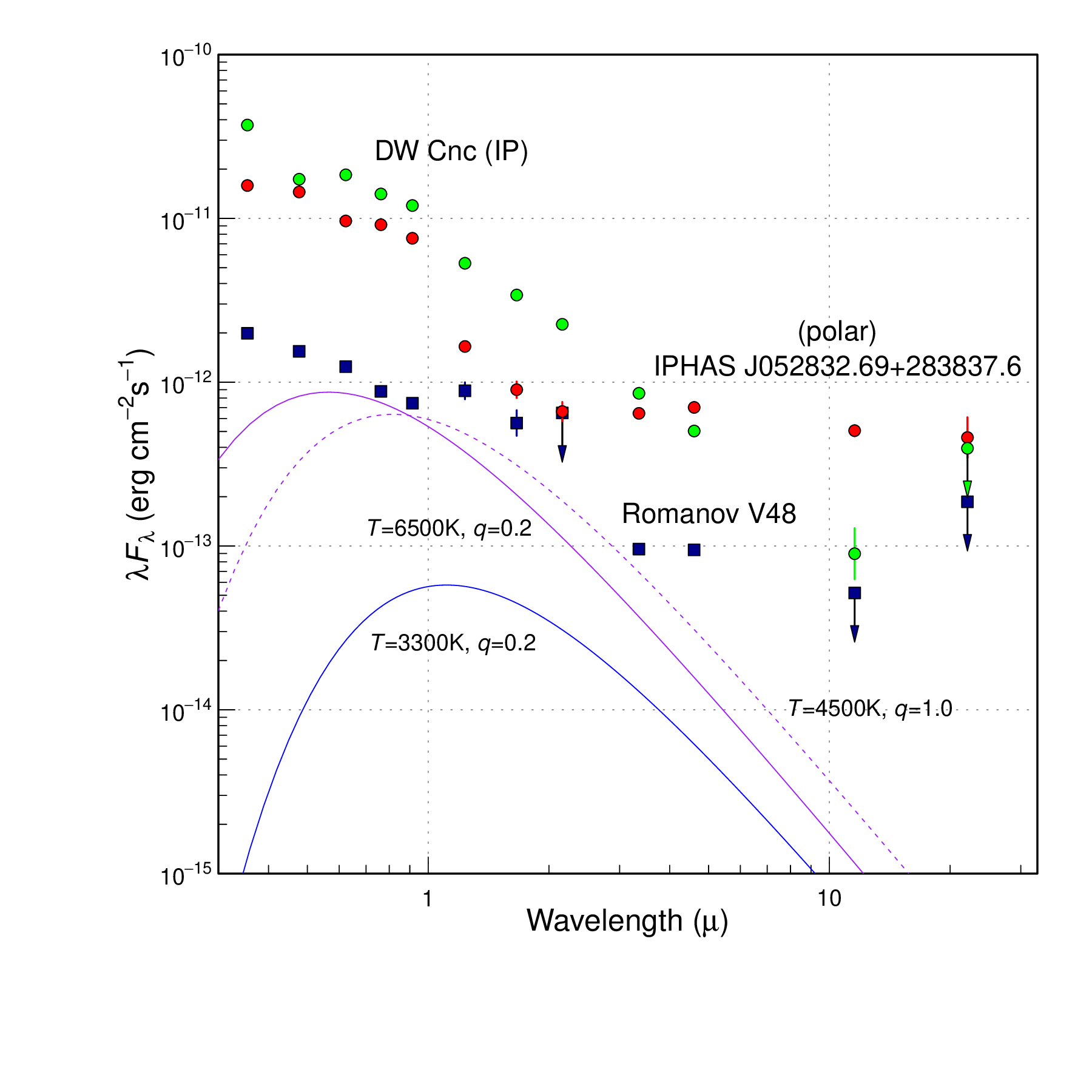}
\caption{The spectral energy distribution (SED) of
  Romanov V48 (filled squares).
  The figure is based on SDSS \citep{SDSS,SDSS16}
  $u'$,$g'$,$r'$,$i'$,$z'$ magnitudes, 2MASS $J$, $H$, $K_{\rm s}$
  magnitudes and WISE W1--W4 magnitudes.  Zero-point calibrations 
  of the 2MASS magnitudes used \citet{coh032MASScalib}. 
  The short-$P_{\rm orb}$ intermediate polar DW Cnc (green circles)
  and the short-$P_{\rm orb}$ polar
  IPHAS J052832.69$+$283837.6 (red circles)
  are shown for a comparison.  The arrows represent upper limits.
  The error bars represent 1$\sigma$ errors arising from
  photometric errors only.
  1$\sigma$ errors arising from the uncertainty in the distance
  of Romanov V48 are 0.20~mag.
  The curves represent the expected black-body radiation for
  the secondary filling the Roche lobe of a binary with
  a period of 0.10231~d at the distance of Romanov V48.
  Three combinations of the mass ratios
  ($q$) and temperatures are shown as a typical M-type star
  and extreme limits.  Any donor filling the Roche lobe
  cannot explain the infrared SED of Romanov V48.
}
\label{fig:sedfig}
\end{center}
\end{figure*}

   Romanov V48 has an excess infrared emission in
2MASS $J$, $H$ \citep{2MASS} and the Wide-field Infrared Survey
Explorer (WISE: \cite{wri10WISE}) W1 and W2 bands
(the object was below the detection limit in 2MASS $K_{\rm s}$,
WISE W3 and W4).
Even a donor with an evolved core
[such as in QZ Ser \citep{tho02qzser}]
cannot explain the spectral energy distribution (SED)
in the infrared, while the infrared SED
of DW Cnc is on the Rayleigh-Jeans tail of a hot object
(figure \ref{fig:sedfig}).
The infrared excess resembles that of a polar as shown in the figure.
IPHAS J052832.69$+$283837.6 was chosen as an example of
a polar \citep{bor16polars} [see e.g. \citet{har15polarWISE}
for infrared SEDs of other polars].
The infrared excess in Romanov V48 may suggest that
the emission mechanism in this system is similar to those
in polars.  The large amplitudes of the spin pulses may also
suggest a magnetic field stronger than the majority of IPs.
Having a large $P_{\rm spin}/P_{\rm orb}$,
Romanov V48 may be an intermediate object between IPs
and polars.  Polarimetric, X-ray and spectroscopic observations
are encouraged.

\section*{Acknowledgements}

This work was supported by JSPS KAKENHI Grant Number 21K03616.
The author is grateful to the ZTF team
for making their data available to the public.
We are grateful to Naoto Kojiguchi for
helping downloading the ZTF data.
This research has made use of the AAVSO Variable Star Index
and NASA's Astrophysics Data System.

Based on observations obtained with the Samuel Oschin 48-inch
Telescope at the Palomar Observatory as part of
the Zwicky Transient Facility project. ZTF is supported by
the National Science Foundation under Grant No. AST-1440341
and a collaboration including Caltech, IPAC, 
the Weizmann Institute for Science, the Oskar Klein Center
at Stockholm University, the University of Maryland,
the University of Washington, Deutsches Elektronen-Synchrotron
and Humboldt University, Los Alamos National Laboratories, 
the TANGO Consortium of Taiwan, the University of 
Wisconsin at Milwaukee, and Lawrence Berkeley National Laboratories.
Operations are conducted by COO, IPAC, and UW.

The ztfquery code was funded by the European Research Council
(ERC) under the European Union's Horizon 2020 research and 
innovation programme (grant agreement n$^{\circ}$759194
-- USNAC, PI: Rigault).

This publication makes use of data products from the Wide-field
Infrared Survey Explorer, which is a joint project of
the University of California, Los Angeles, and the
Jet Propulsion Laboratory/California Institute of Technology,
funded by the National Aeronautics and Space Administration. 

This publication makes use of data products from the
Two Micron All Sky Survey, which is a joint project of the
University of Massachusetts and the Infrared Processing and
Analysis Center/California Institute of Technology, funded by
the National Aeronautics and Space Administration and the
National Science Foundation.

\section*{List of objects in this paper}
\xxinput{objlist.inc}

\section*{References}

We provide two forms of the references section (for ADS
and as published) so that the references can be easily
incorporated into ADS.

\renewcommand\refname{\textbf{References (for ADS)}}

\newcommand{\noop}[1]{}\newcommand{\hyphalt}{-}

\xxinput{v48aph.bbl}

\renewcommand\refname{\textbf{References (as published)}}

\xxinput{v48.bbl.vsolj}


\begin{thebibliography}{}

\bibitem[{Ahumada} et~al.(2020)]{SDSS16}
  {Ahumada}, R., {et~al.}\ 2020, ApJS, 249, 3 (arXiv:1912.02905)

\bibitem[{Araujo-Betancor} et~al.(2005)]{ara05v455and}
  {Araujo-Betancor}, S., {et~al.}\ 2005, A\&A, 430, 629
  (arXiv:astro-ph/0410223)

\bibitem[{Baran} et~al.(2021)]{bar21tesssdBs}
  {Baran}, A.~S., {Sahoo}, S.~K., {Sanjayan}, S., \& {Ostrowski}, J.\ 2021,
  MNRAS, 503, 3828 (arXiv:2105.01077)

\bibitem[{Borisov} et~al.(2016)]{bor16polars}
  {Borisov}, N.~V., {Gabdeev}, M.~M., \& {Afanasiev}, V.~L.\ 2016, Astrophys.\
  Bull., 71, 95 (https://doi.org/10.1134/S1990341316010107)

\bibitem[{Buckley} et~al.(1998)]{buc98v1025cen}
  {Buckley}, D. A.~H., {Cropper}, M., {Ramsay}, G., \& {Wickramasinghe}, D.~T.\
  1998, MNRAS, 299, 83 (https://doi.org/10.1046/j.1365-8711.1998.01744.x)

\bibitem[{Butters} et~al.(2008)]{but08v515and}
  {Butters}, O.~W., {Norton}, A.~J., {Hakala}, P., {Mukai}, K., \& {Barlow},
  E.~J.\ 2008, A\&A, 487, 271 (arXiv:0806.0751)

\bibitem[{Chambers} et~al.(2016)]{PS1}
  {Chambers}, K.~C., {et~al.}\ 2016, arXiv e-prints, arXiv:1612.05560

\bibitem[{Cohen} et~al.(2003)]{coh032MASScalib}
  {Cohen}, M., {Wheaton}, W.~A., \& {Megeath}, S.~T.\ 2003, AJ, 126, 1090
  (arXiv:astro-ph/0304350)

\bibitem[{Cutri} et~al.(2003)]{2MASS}
  {Cutri}, R.~M., {et~al.}\ 2003, {2MASS} {All Sky Catalog} of point sources
 (NASA/IPAC Infrared Science Archive)

\bibitem[{Fernie}(1989)]{fer89error}
  {Fernie}, J.~D.\ 1989, PASP, 101, 225 (https://doi.org/10.1086/132426)

\bibitem[{Gaia Collaboration} et~al.(2018)]{GaiaDR2}
  {Gaia Collaboration}, {et~al.}\ 2018, A\&A, 616, A1 (arXiv:1804.09365)

\bibitem[{Gaia Collaboration} et~al.(2021)]{GaiaEDR3}
  {Gaia Collaboration}, {et~al.}\ 2021, A\&A, 649, A1 (arXiv:2012.01533)

\bibitem[{G{\"a}nsicke} et~al.(2009)]{gan09SDSSCVs}
  {G{\"a}nsicke}, B.~T., {et~al.}\ 2009, MNRAS, 397, 2170 (arXiv:0905.3476)

\bibitem[{Halpern} and {Thorstensen}(2015)]{hal15SwiftCVs}
  {Halpern}, J.~P., \& {Thorstensen}, J.~R.\ 2015, AJ, 150, 170
  (arXiv:1510.00703)

\bibitem[{Harrison} and {Campbell}(2015)]{har15polarWISE}
  {Harrison}, T.~E., \& {Campbell}, R.~K.\ 2015, ApJS, 219, 32
  (https://doi.org/10.1088/0067-0049/219/2/32)

\bibitem[{Heinze} et~al.(2018)]{hei18ATLASvar}
  {Heinze}, A.~N., {et~al.}\ 2018, AJ, 156, 241
  (https://doi.org/10.3847/1538-3881/aae47f)

\bibitem[{Hellier} et~al.(2002)]{hel02v1025cen}
  {Hellier}, C., {Wynn}, G.~A., \& {Buckley}, D. A.~H.\ 2002, MNRAS, 333, 84
  (arXiv:astro-ph/0201474)

\bibitem[{Howell} et~al.(1997)]{how97periodminimum}
  {Howell}, S.~B., {Rappaport}, S., \& {Politano}, M.\ 1997, MNRAS, 287, 929
  (https://doi.org/10.1093/mnras/287.4.929)

\bibitem[{Jablonski} and {Busko}(1985)]{jab85exhya}
  {Jablonski}, F., \& {Busko}, I.~C.\ 1985, MNRAS, 214, 219
  (https://doi.org/10.1093/mnras/214.2.219)

\bibitem[{Kato} et~al.(2015)]{kat15ccscl}
  {Kato}, T., {Hambsch}, F.-J., {Oksanen}, A., {Starr}, P., \& {Henden}, A.\
  2015, PASJ, 67, 3 (arXiv:1409.8004)

\bibitem[{Kato}(2022)]{kat22stageA}
  {Kato}, T.\ 2022, VSOLJ\ Variable\ Star\ Bull., 89, (arXiv:2201.02945)

\bibitem[{Kato} et~al.(2010)]{Pdot2}
  {Kato}, T., {et~al.}\ 2010, PASJ, 62, 1525 (arXiv:1009.5444)

\bibitem[{Kemp} et~al.(2002)]{kem02htcam}
  {Kemp}, J., {Patterson}, J., {Thorstensen}, J.~R., {Fried}, R.~E.,
  {Skillman}, D.~R., \& {Billings}, G.\ 2002, PASP, 114, 623
  (arXiv:astro-ph/0204227)

\bibitem[{Knigge} et~al.(2011)]{kni11CVdonor}
  {Knigge}, C., {Baraffe}, I., \& {Patterson}, J.\ 2011, ApJS, 194, 28
  (arXiv:1102.2440)

\bibitem[{Knigge}(2006)]{kni06CVsecondary}
  {Knigge}, C.\ 2006, MNRAS, 373, 484 (arXiv:astro-ph/0609671)

\bibitem[{Kolb}(1993)]{kol93CVpopulation}
  {Kolb}, U.\ 1993, A\&A, 271, 149

\bibitem[{Kozhevnikov}(2012)]{koz12v515and}
  {Kozhevnikov}, V.~P.\ 2012, MNRAS, 422, 1518 (arXiv:1202.2493)

\bibitem[{Martin} et~al.(2005)]{GALEX}
  {Martin}, D.~C., {et~al.}\ 2005, ApJ, 619, L1 (arXiv:astro-ph/0411302)

\bibitem[{Masci} et~al.(2019)]{ZTF}
  {Masci}, F.-J., {et~al.}\ 2019, PASP, 131, 018003 (arXiv:1902.01872)

\bibitem[{Mowlavi} et~al.(2021)]{mow21Gaiavar}
  {Mowlavi}, N., {et~al.}\ 2021, A\&A, 648, A44 (arXiv:2009.07746)

\bibitem[{Norton} et~al.(2008)]{nor08IPaccretion}
  {Norton}, A.~J., {Butters}, O.~W., {Parker}, T.~L., \& {Wynn}, G.~A.\ 2008,
  ApJ, 672, 524 (arXiv:0709.4186)

\bibitem[{Ofek} et~al.(2020)]{ofe20ZTFvar}
  {Ofek}, E.~O., {Soumagnac}, M., {Nir}, G., {Gal-Yam}, A., {Nugent}, P.,
  {Masci}, F., \& {Kulkarni}, S.~R.\ 2020, MNRAS, 499, 5782 (arXiv:2007.01537)

\bibitem[{Patterson} et~al.(2004)]{pat04dwcnc}
  {Patterson}, J., {et~al.}\ 2004, PASP, 116, 516 (arXiv:astro-ph/0404082)

\bibitem[{Patterson}(1994)]{pat94IPreview}
  {Patterson}, J.\ 1994, PASP, 106, 209 (https://doi.org/10.1086/133375)

\bibitem[{Scargle}(1982)]{LombScargle}
  {Scargle}, J.~D.\ 1982, ApJ, 263, 835 (https://doi.org/10.1086/160554)

\bibitem[{Stellingwerf}(1978)]{PDM}
  {Stellingwerf}, R.~F.\ 1978, ApJ, 224, 953 (https://doi.org/10.1086/156444)

\bibitem[{Sterken} et~al.(1983)]{ste83exhya}
  {Sterken}, C., {Vogt}, N., {Freeth}, R., {Kennedy}, H.~D., {Marino}, B.~F.,
  {Page}, A.~A., \& {Walker}, W.~S.~G.\ 1983, A\&A, 118, 325

\bibitem[{Thorstensen} et~al.(2002)]{tho02qzser}
  {Thorstensen}, J.~R., {Fenton}, W.~H., {Patterson}, J.~O., {Kemp}, J.,
  {Halpern}, J., \& {Baraffe}, I.\ 2002, PASP, 114, 1117
  (arXiv:astro-ph/0206435)

\bibitem[{Thorstensen} and {Halpern}(2013)]{tho13XrayCVs}
  {Thorstensen}, J.~R., \& {Halpern}, J.\ 2013, AJ, 146, 107 (arXiv:1308.5016)

\bibitem[{Tonry} et~al.(2018)]{ATLAS}
  {Tonry}, J.~L., {et~al.}\ 2018, PASP, 130, 064505 (arXiv:1802.00879)

\bibitem[{Tovmassian} et~al.(1998)]{tov98htcam}
  {Tovmassian}, G.~H., {et~al.}\ 1998, A\&A, 335, 227 (arXiv:astro-ph/9803234)

\bibitem[{Watson} et~al.(2006)]{wat06VSX}
  {Watson}, C.~L., {Henden}, A.~A., \& {Price}, A.\ 2006, Society\ for\
  Astronom.\ Sciences\ Ann.\ Symp., 25, 47

\bibitem[{Woudt} et~al.(2012)]{wou12ccscl}
  {Woudt}, P.~A., {et~al.}\ 2012, MNRAS, 427, 1004 (arXiv:1208.5936)

\bibitem[{Wright} et~al.(2010)]{wri10WISE}
  {Wright}, E.~L., {et~al.}\ 2010, AJ, 140, 1868 (arXiv:1008.0031)

\bibitem[{York} et~al.(2000)]{SDSS}
  {York}, D.~G., {et~al.}\ 2000, AJ, 120, 1579 (arXiv:astro-ph/0006396)

\end{thebibliography}
\end{document}